\documentstyle[12pt]{article}

\newcommand{\EQ}{\begin{equation}}
\newcommand{\EN}{\end{equation}}

\def \prl {Phys. Rev. Lett. }
\def \plb {Phys. Lett. B }
\def \npb {Nucl. Phys. B }
\def \prd {Phys. Rev. D }

\newcommand{\bea}{\begin{eqnarray}}
\newcommand{\eea}{\end{eqnarray}}
\newcommand{\bean}{\begin{eqnarray*}}
\newcommand{\eean}{\end{eqnarray*}}

\begin{document}
\topmargin 0pt
\oddsidemargin=-0.4truecm
\evensidemargin=-0.4truecm
\renewcommand{\thefootnote}{\fnsymbol{footnote}}
\setcounter{page}{1}
\begin{titlepage}
\begin{flushright}
FTUV/95-24 IFIC/95-24 \\
\end{flushright}

\vspace{0.7cm}
\begin{center}
{\Large
What can we learn from the $Z \rightarrow b\bar{b} $ vertex?
}
\vspace{1.2cm}

{\Large Denis Comelli} \footnote{E-mail: Comelli@evalvx.ific.uv.es
   supported by the "Ministerio de la scientia y de la instruccion  de
Espana"}\\
\vspace{0.2cm}
{\em University of Valencia 46100 Burjassot (Valencia),
Spain}\\

\vspace{0.4cm}

\end{center}
\begin{abstract}
I stress the fact
that a complete study of possible New Physics effects in the
$Z \rightarrow b\bar{b}$ vertex
requires a combined analysis of the ratio
$R_b= \Gamma(Z \rightarrow b\bar{b})/ \Gamma(Z \rightarrow hadrons)$
and the longitudinally polarized forward backward asymmetry, $A_b$.

This is illustrated with a number of models such as the two Higgs-doublet
model, the MSSM, Technicolour and models with an extra $Z'$.
The use of effective lagrangean techniques is briefly discussed.
\vspace{0.4cm}

\end{abstract}

\end{titlepage}
\renewcommand{\thefootnote}{\arabic{footnote}}
\setcounter{footnote}{0}
\newpage
{\bf \large }

The remarkable experimental effort done by the LEP collaboration has
brought some hadronic processes into the realm of high precision
measurements.
The best example arises in the ratio
$R_b= \Gamma(Z \rightarrow b\bar{b})/ \Gamma(Z \rightarrow hadrons)$
which is now known with an error of the order of 1$\%$
($R_b= 0.2204 \pm 0.0020$).
This quantity is a direct measurement of the non oblique vertex
correction, since the oblique and QCD corrections cancel in the ratio.

The effective vertex $Z \rightarrow b\bar{b}$ can be parametrized in terms of
two form factors, the left handed $g_L$ and the right handed $g_R$:
\EQ
\frac{g}{2 \cos \theta_W} Z^{\mu}  \bar{\psi}_b
(\gamma_{\mu}(g_R\gamma_R +g_L\gamma_L ))\psi_b\ ,
\EN
where $\gamma_{R,L}=(1\pm \gamma_5)$. A third form factor,
the so called magnetic moment, is neglected.

The effective couplings may be written as
$g_{L,R}= g_{L,R}^0 +\delta g_{L,R}$
with
$g_{L}^0=- 1/2+ 1/3
\sin^2 \theta_W $ and $ g_{L}^0=1/3
\sin^2 \theta_W $.
The shifts $\delta g_{L,R}$ contain the
effect of the non oblique  Radiative Corrections (RC) of the Standard Model
(SM), due to the would be goldstone boson exchange
and QCD effects due to the non negligible $b$ mass,
as well as
possible New Physics (NP) effects.
In order to determine the two form factors separately we need two
independent measurements of the same vertex.
One is provided by $R_b$ and the other by the longitudinally polarized
forward backward asymmetry defined as :
\EQ
A_b= \frac{
\sigma(e_L\rightarrow b_F)-\sigma(e_R \rightarrow b_F)-\sigma(e_L \rightarrow
b_B)+\sigma(e_R \rightarrow b_B)}
{\sigma(e_L \rightarrow b_F)+\sigma(e_R \rightarrow b_F)+\sigma(e_L \rightarrow
b_B)+\sigma(e_R \rightarrow b_B)}\ ,
\EN
which is directly measured at SLAC \cite{slac} as well as in the
forward backward asymmetry at LEP \cite{lep}.
Remarkably  $A_b$  depends  only on the $Z$
coupling of the $b$ quark  \cite{blondel}.
\EQ
A_b= \frac{g_L^2-g_R^2}{g_L^2+g_R^2}
\EN

The effect of  the vertex corrections
can be outlined through the ratios  of
the $Z\rightarrow b \bar{b}$ rate and asymmetry to those
of the decay $Z \rightarrow s \bar{s}$,
\EQ
\frac{\Gamma_b}{\Gamma_s}=1+\delta_{bV} \;\; {\rm and}
\;\; \frac{A_b}{A_s}=1+\eta_b
\EN
We may write
$\delta_{bV}$ and $\eta_b$
in terms of the shift $\delta g_{L,R}$ as,
\EQ
\delta_{bV}= -\frac{4(1+b)}{(1+b^2)} [ \delta g_L
-\frac{(1-b)}{(1+b)}\delta g_R]
\EN
\EQ
\eta_b=-\frac{4(1-b)(1+b)^2}{b(1+b^2)}[\delta
g_R+\frac{(1-b)}{(1+b)}\delta g_L]\ ,
\EN
where $b=1-4/3 \sin^2 \theta_W$.
Thus,
$\delta_{bV}$ and $\eta_b$
are naturally orthogonal in the ($\delta g_L,\delta g_R$) basis,
so that the effects that would not contribute to one observable will
be revealed by the other one and vice-versa.
In particular,
the NP effects contributing only to the left handed coupling
($\delta g_R=0$) imply the correlation $ \eta_b/\delta_{bV}=0.068$.
Conversely,
for the physics that contributes only to the right handed
($\delta g_L=0$) one finds $ \eta_b/\delta_{bV}=-2.068$.
This means that $ \eta_b$ is very
sensitive to $\delta g_R$ whereas $\delta_{bV}$ is very sensitive
to $\delta g_L$.

The leading SM corrections to the $Z \rightarrow b\bar{b}$ vertex are
very important.
They are due  to  the exchange of the would be
Goldstone boson in the t'Hooft-Feynman gauge,
and,
in the limit of large top mass ($m_t>>m_W$),
are quadratic in $m_t$ contributing only to
$\delta g_L$ \cite{akundov}.
As a consequence  any  deviation of $ \eta_b$ from zero is a signal of
NP.

Using the present  value  of $m_t=174 \pm20$ GeV \cite{fermilab},
the LEP data  indicate that the SM prediction for $R_b$ might be
inconsistent with such a heavy top up to 2.5 $\sigma$,
and that NP contribution, $\delta_{bV}^{NP}$,
must be positive.
This initial hint seems very interesting and justifies a thorough
investigation of possible NP models providing $\delta_{bV}^{NP}>0$,
together with an analysis of their predictions for $\eta_b$.

\bigskip

$\Box$   The simplest extension of the SM is the Two Higgs Doublet
Model (THDM) in which the electroweak symmetry breaking sector
involves two fundamental scalar doublets $H_{1,2}$.
The physical spectrum contains an extra pair of charged scalar
particles $H^{\pm}$ , a pseudoscalar $A^0$,
and two neutral CP even scalars $h^0$ and $H^0$.

The parameter $\tan \beta$, defined as the ratio of the two vevs
$<H_1>/<H_2>$, allows us to distinguish two different regions for the RC
\cite{two}:

1) $\tan \beta \sim 1$: for which the only important contributions
come from the $H^{\pm}$ loops. One finds $\delta g_R=0$, $\delta g_L \sim
0(m_t^2/M_H^2)$ and consequently
$\delta_{bV} <0$ and $\eta_b\simeq 0$ , in contradiction with the previous
expectation.

2) $\tan \beta \geq m_t/m_b$ : in this region the Yukawa coupling of
the $b$ quark can be comparable to that of the $t$ quark so
that the process involving intermediate $b$ quark and neutral scalars
become important.
For the charged sector we find $\delta g_L =0$, $\delta g_R=0(m_b^2
\tan ^2\beta/M_H^2)$  with $\delta_{bV} >0$ and $\eta_b\simeq
-2\delta_{bV}$.
For the neutral sector $\delta g_R$ and $\delta g_L$ are $0(m_b^2
\tan ^2\beta/M_H^2)$ with $\delta_{bV} >0$ and $\eta_b\geq 0$.
In particular, for a light
pseudoscalar $A^0$,
these last contributions can be very large.

\bigskip

$\Box$ The next models that I discuss are the SUSY theories .
The Higgs sector of the Minimal SUSY
model coincides with that of THDM,
once the mass relations required by SUSY are imposed.
In addition to these contributions discussed above,
we also have contributions coming from intermediate states of the
superpartners of the SM particles .
For light $M_{A^0}$ and large $\tan \beta$ the effect can be important
and goes in the right direction, namely, $\delta_{bV} >0$
 and $\eta_b\geq 0$.\cite{two}.
These superparticle contributions introduce a lot of new free
parameters (the masses of the stop-sbottom left and right and the
respective mixing angles, the $\mu$ term, and the gaugino
masses)\cite{susy}.

The contribution from the strong interactions involving the gluino are
practically negligible.
Sizeable corrections can come instead from the stop-chargino sector
with

$\delta g_L\simeq 0(m_t^2/(\sin^2 \beta M_{SUSY}^2))$  and
$\delta_{bV} >0$
for small $\tan \beta$; and

$\delta g_R
\simeq 0(m_b^2 \tan^2\beta/ M_{SUSY}^2)$
for large $\tan \beta$ ,
with $\delta_{bV}\simeq-1/2 \eta_b<0$.

For large $\tan \beta$ there are also  sizeable
contributions from the neutral
fermionic supersector (neutralino-sbottom loop)
which act along the right experimental direction ,

 $\delta g_L\simeq
-\delta g_R\simeq 0(m_b^2 \tan^2\beta/ M_{SUSY}^2)$
with $\eta_b\simeq-0.3 \delta_{bV}<0$.

In general it is possible to state that, if the
discrepancy of $R_b$ is to be explained by the MSSM,
then direct observation of superpartners at LEPII or
at an upgraded Tevatron can be expected
\cite{kane}.

\bigskip

$\Box$  These vertex corrections come out  to be
 very sensitive also to completely
different models, like Technicolor, in which the dynamical electroweak
symmetry breaking is induced by the vev of bilinear fermion (instead of a
fundamental scalar particle).

To provide simultaneously the masses of the W and Z bosons and of the
ordinary fermions, additional gauge interactions (Extended
Technicolor (ETC)) are  introduced.

The effective low energy theory provides a sizeable coupling between
the third quark family and the Technifermions inducing a big shift in
$\delta g_L\sim O(m_t/4\pi v)$.
This
implies a negative contribution to $\delta_{bV}$,
which is about 8 $\sigma$
away from the experimental value \cite{technicolor}.
Such a negative result, added to  the presence of FCNC, puts
this models in serious trouble.

Also, in the more popular walking technicolor the effect can be at most
a factor 2 smaller \cite{etc}.
\bigskip

$\Box$  Another interesting class of models that can affect this
vertex are theories containing extra gauge bosons.
When this extra bosons mix with the SM one, the rotation to the mass
basis induces a change in the coupling of the physical Z to the
fermions which is proportional to $\delta g_{LR}^{f}\simeq -x
m_Z^2/M_{Z'}^2 g_{Z' LR}^{f}  $, where $x$ is a model dependent
quantity and $g_{Z' LR}^{f}$ are the coupling of the extra $Z'$ with
the $f$-fermions.

The limits on the mixing angle
of the extra gauge bosons,
$\theta= x m_Z^2/M_{Z'}^2 $,
coupling universally to the fermions
(as in Superstring inspired and Gut models)
are so strong that the mixing effect on the
$Z\rightarrow b \bar{b}$ vertex cannot exceed the 1 $\%$.

Recently,
some Technicolor inspired models have been explored
in which the extra $Z'$ responsible for the generation of the top quark mass
couples more strongly to the $b_L$, $t_L$ and $t_R$
than to the other fermions.
The implications of these models are studied in ref.\cite{holdom}  in which
it is found that the existence of contributions proportional to $\delta
g_L=-\delta g_R\;\;(\eta_b/\delta_{bV}\simeq -0.3)$  open a possible
reconciliation of the ETC with the experiments.

\bigskip

$\Box$  Finally, I would like to discuss
an alternative method to investigate the
effects of possible NP on the $Zb \bar{b}$ vertex.
I will mention only the results of a general investigation
that can be found in ref \cite{verza},
which exploits effective lagrangean techniques.
In this approach the Lagrangean is written as
$L=L_{SM}+ \sum_i f_i O_i/\Lambda^2$,
where $L_{SM}$ is the SM Lagrangean,
$\Lambda$ is the scale at which NP appears and $ O_i$ are all the
gauge and CP invariant
operators of dimension 6 which contribute to the RC.

Using the ansatz that the NP generates only operators containing gauge
boson fields (11 independent operators) and operators  with
at least one $t_R$ field (the $t_R$ field is related to the $m_t$ mass)
(14 operators) they find that the leading order RC to
$Zb \bar{b}$ [of order $0(m_t^2/\Lambda^2 \log \Lambda^2$)]
obey the following pattern:

1)only two $O_i$ of the gauge sector
and four of the fermionic-$t_R$ sector affect
$\delta g_L$, thus yielding $\eta_b\simeq 0$;

2)only one operator with one $t_R$ field affects $\delta g_R$
and, hence, affecting also $\eta_b$.

In practice in this formalism a sizeable non zero value of $\eta_b$
is directly related to the presence of one operator $O_{t_R}$.
However, this technique must be used with care, due to the possible
presence of large finite contributions of order O($m_t^2/\Lambda^2$)
\cite{zheng}.

\vspace{1cm}

{\bf In conclusion}
I have discussed the possible "directions" of NP on the
plane ($\delta_{bV}, \eta_b$)  .

In the first quadrant we find the contributions of the THDM,
of most part of the parameter space of SUSY models,
and of modest Z' contributions;
in the fourth quadrant there are instead the extra Z' models
inspired by Technicolor and the presence of effects from the $O_{t_R}$
operator.

The next improved measurements of the forward backward asymmetry as
well as $A_b$ in SLAC  can be of help to disentangle indications of
NP.

\vspace{0.5cm}

I would like to take this opportunity to thank the organizers for
the excellent environment and interesting discussions they
have extended to us at this conference.
I would also like to thank A.\ Rossi and J.P Silva for reading the manuscript.

\end{document}